\documentclass[11pt]{article}
\usepackage{amssymb}

\usepackage{graphicx}
\usepackage{amsmath}
\usepackage{makeidx}
\usepackage{indentfirst}

\newcounter{resultnum}[section]

\newcounter{conclusionnum}[section]

\newcounter{conditionnum}[section]

\newcounter{conjecturenum}[section]

\newcounter{examplenum}[section]

\newcounter{exercisenum}[section]

\newcounter{lemmanum}[section]

\newcounter{notationnum}[section]
\newtheorem{theorem}{Theorem}[section]

\newcounter{theoremnum}[section]

\newcounter{definitionnum}[section]

\newcounter{corollarynum}[section]

\newcounter{remarknum}[section]

\newcounter{propositionnum}[section]

\newcounter{acknowledgementnum}[section]

\newcounter{algorithmnum}[section]

\newcounter{axiomnum}[section]

\newcounter{casenum}[section]

\newcounter{claimnum}[section]

\newcounter{summarynum}[section]

\newcounter{problemnum}[section]

\begin{document}

\title{Fractional Exact Solutions\\
and Solitons in Gravity}
\date{July 16, 2010}
\author{\textbf{Dumitru Baleanu}\thanks{%
On leave of absence from Institute of Space Sciences, P. O. Box, MG-23, R
76900, Magurele--Bucharest, Romania;\ dumitru@cancaya.edu.tr,
baleanu@venus.nipne.ro} \\
\textsl{\small Department of Mathematics and Computer Sciences,} \\
\textsl{\small \c Cankaya University, 06530, Ankara, Turkey } \\
\and 
\textbf{Sergiu I. Vacaru} \thanks{%
sergiu.vacaru@uaic.ro;\ http://www.scribd.com/people/view/1455460-sergiu}
\and \textsl{\small Science Department, University "Al. I. Cuza" Ia\c si, }
\\
\textsl{\small 54, Lascar Catargi street, Ia\c si, Romania, 700107}}
\maketitle

\begin{abstract}
We survay our recent results on fractional gravity theory. It is also
provided the Main Theorem on encoding of geometric data (metrics and
connections in gravity and geometric mechanics) into solitonic hierarchies.
Our approach is based on Caputo fractional derivative and nonlinear
connection formalism.

\vskip5pt


\vskip3pt


PACS2010:\ 02.30.Ik, 45.10Hj, 05.45.Yv, 02.40.Yy, 45.20.Jj, 04.50.Kd
\end{abstract}

\tableofcontents

\section{Introduction}

Recently, we extended the fractional calculus to Ricci flow theory, gravity
and geometric mechanics, solitonic hierarchies etc \cite%
{vrfrf,vrfrg,bv1,bv2,bv3,bv4}. In this work, we outline some basic geometric
constructions related to fractional derivatives and integrals and their
applications in modern physics and mechanics.

Our approach is also connected to a method when nonholonomic deformations of
geometric structures\footnote{%
determined by a fundamental Lagrange/ Finsler / Hamilton generating function
(or, for instance, and Einstein metric)} induce a canonical connection,
adapted to a necessary type nonlinear connection structure, for which the
matrix coefficients of curvature are constant \cite{vacap,vanco}. For such
an auxiliary connection, it is possible to define a bi--Hamiltonian
structure and derive the corresponding solitonic hierarchy.

The paper is organized as follows: In section 2, we outline the geometry of
N--adapted fractional manifolds and provide an introduction to fractional
gravity. In section 3, we show how fractional gravitational field equations
can be solved in a general form. Section 4 is devoted to the Main Theorem on
fractional solitonic hierarchies corresponding to metrics and connections in
fractional gravity. The Appendix contains necessary definitions and formulas
on Caputo fractional derivatives.

\vskip5pt \textbf{Acknowledgement: } This paper summarizes the results presented in 
our talk at Conference ''New Trends in Nanotechnology and Nonlinear
Dynamical Systems'', 25--27 July, 2010, \c{C}ankaya University, Ankara,
Turkey.

\section{Fractional Nonholonomic Manifolds and Gravity}

Let us consider a ''prime'' nonholonomic manifold $\mathbf{V}$ is of integer
dimension $\dim $ $\mathbf{V}=n+m,n\geq 2,m\geq 1.$\footnote{%
A nonholonomic manifold is a manifold endowed with a non--integrable
(equivalently, nonholonomic, or anholonomic) distribution.}\ Its fractional
extension $\overset{\alpha }{\mathbf{V}}$ is modelled by a quadruple $(%
\mathbf{V},\overset{\alpha }{\mathbf{N}},\overset{\alpha }{\mathbf{d}},%
\overset{\alpha }{\mathbf{I}}),$ where $\overset{\alpha }{\mathbf{N}}$ is a
nonholonomic distribution stating a nonlinear connection (N--connection)
structure (for details, see Appendix \ref{sappa} with explanations for
formula (\ref{whit}). The fractional differential structure $\overset{\alpha
}{\mathbf{d}}$ is determined by Caputo fractional derivative (\ref{lfcd})
following formulas (\ref{frlcb}) and (\ref{frlccb}). The non--integer
integral structure $\overset{\alpha }{\mathbf{I}}$ is defined by rules of
type (\ref{aux01}).

A nonlinear connection (N--connection) $\overset{\alpha }{\mathbf{N}}$ \ for
a fractional space $\overset{\alpha }{\mathbf{V}}$ is defined by a
nonholonomic distribution (Whitney sum) with conventional h-- and
v--subspaces, $\underline{h}\overset{\alpha }{\mathbf{V}}$ and $\underline{v}%
\overset{\alpha }{\mathbf{V}},$%
\begin{equation}
\overset{\alpha }{\underline{T}}\overset{\alpha }{\mathbf{V}}=\underline{h}%
\overset{\alpha }{\mathbf{V}}\mathbf{\oplus }\underline{v}\overset{\alpha }{%
\mathbf{V}}.  \label{whit}
\end{equation}

A fractional N--connection is defined by its local coefficients $\overset{%
\alpha }{\mathbf{N}}\mathbf{=}\{\ ^{\alpha }N_{i}^{a}\},$ when
\begin{equation*}
\overset{\alpha }{\mathbf{N}}\mathbf{=}\ ^{\alpha
}N_{i}^{a}(u)(dx^{i})^{\alpha }\otimes \overset{\alpha }{\underline{\partial
}}_{a}.
\end{equation*}%
For a N--connection $\overset{\alpha }{\mathbf{N}},$ we can always construct
a class of fractional (co) frames (N--adapted) linearly depending on $\
^{\alpha }N_{i}^{a},$
\begin{eqnarray}
\ ^{\alpha }\mathbf{e}_{\beta } &=&\left[ \ ^{\alpha }\mathbf{e}_{j}=\overset%
{\alpha }{\underline{\partial }}_{j}-\ ^{\alpha }N_{j}^{a}\overset{\alpha }{%
\underline{\partial }}_{a},\ ^{\alpha }e_{b}=\overset{\alpha }{\underline{%
\partial }}_{b}\right] ,  \label{dder} \\
\ ^{\alpha }\mathbf{e}^{\beta } &=&[\ ^{\alpha }e^{j}=(dx^{j})^{\alpha },\
^{\alpha }\mathbf{e}^{b}=(dy^{b})^{\alpha }+\ ^{\alpha
}N_{k}^{b}(dx^{k})^{\alpha }].  \label{ddif}
\end{eqnarray}%
The nontrivial nonholonomy coefficients are computed $\ ^{\alpha }W_{ib}^{a}=%
\overset{\alpha }{\underline{\partial }}_{b}\ ^{\alpha }N_{i}^{a}$ and $\
^{\alpha }W_{ij}^{a}=\ ^{\alpha }\Omega _{ji}^{a}=\ ^{\alpha }\mathbf{e}%
_{i}\ ^{\alpha }N_{j}^{a}-\ ^{\alpha }\mathbf{e}_{j}\ ^{\alpha }N_{i}^{a}$
for
\begin{equation*}
\left[ \ ^{\alpha }\mathbf{e}_{\alpha },\ ^{\alpha }\mathbf{e}_{\beta }%
\right] =\ ^{\alpha }\mathbf{e}_{\alpha }\ ^{\alpha }\mathbf{e}_{\beta }-\
^{\alpha }\mathbf{e}_{\beta }\ ^{\alpha }\mathbf{e}_{\alpha }=\ ^{\alpha
}W_{\alpha \beta }^{\gamma }\ ^{\alpha }\mathbf{e}_{\gamma }.
\end{equation*}%
In above formulas, the values $\ ^{\alpha }\Omega _{ji}^{a}$ are called the
coefficients of N--connection curvature. A nonholonomic manifold defined by
a structure $\overset{\alpha }{\mathbf{N}}$ is called, in brief, a
N--anholonomic fractional manifold.

We write a metric structure $\ \overset{\alpha }{\mathbf{g}}=\{\ ^{\alpha
}g_{\underline{\alpha }\underline{\beta }}\}$ on $\overset{\alpha }{\mathbf{V%
}}$ \ in the form
\begin{eqnarray}
\ \overset{\alpha }{\mathbf{g}} &=&\ ^{\alpha }g_{kj}(x,y)\ ^{\alpha
}e^{k}\otimes \ ^{\alpha }e^{j}+\ ^{\alpha }g_{cb}(x,y)\ ^{\alpha }\mathbf{e}%
^{c}\otimes \ ^{\alpha }\mathbf{e}^{b}  \label{m1} \\
&=&\eta _{k^{\prime }j^{\prime }}\ ^{\alpha }e^{k^{\prime }}\otimes \
^{\alpha }e^{j^{\prime }}+\eta _{c^{\prime }b^{\prime }}\ ^{\alpha }\mathbf{e%
}^{c^{\prime }}\otimes \ ^{\alpha }\mathbf{e}^{b^{\prime }},  \notag
\end{eqnarray}%
where matrices $\eta _{k^{\prime }j^{\prime }}=diag[\pm 1,\pm 1,...,\pm 1]$
and $\eta _{a^{\prime }b^{\prime }}=diag[\pm 1,\pm 1,...,\pm 1],$ for the
signature of a ''prime'' spacetime $\mathbf{V,}$ are obtained by frame
transforms $\eta _{k^{\prime }j^{\prime }}=e_{\ k^{\prime }}^{k}\ e_{\
j^{\prime }}^{j}\ _{\ }^{\alpha }g_{kj}$ and $\eta _{a^{\prime }b^{\prime
}}=e_{\ a^{\prime }}^{a}\ e_{\ b^{\prime }}^{b}\ _{\ }^{\alpha }g_{ab}.$

A distinguished connection (d--connection) $\overset{\alpha }{\mathbf{D}}$
on $\overset{\alpha }{\mathbf{V}}$ is defined as a linear connection
preserving under parallel transports the Whitney sum (\ref{whit}). We can
associate a N--adapted differential 1--form
\begin{equation}
\ ^{\alpha }\mathbf{\Gamma }_{\ \beta }^{\tau }=\ ^{\alpha }\mathbf{\Gamma }%
_{\ \beta \gamma }^{\tau }\ ^{\alpha }\mathbf{e}^{\gamma },  \label{fdcf}
\end{equation}%
parametrizing the coefficients (with respect to (\ref{ddif}) and (\ref{dder}%
)) in the form $\ ^{\alpha }\mathbf{\Gamma }_{\ \tau \beta }^{\gamma
}=\left( \ ^{\alpha }L_{jk}^{i},\ ^{\alpha }L_{bk}^{a},\ ^{\alpha
}C_{jc}^{i},\ ^{\alpha }C_{bc}^{a}\right) .$

The absolute fractional differential $\ ^{\alpha }\mathbf{d}=\ _{\ _{1}x}%
\overset{\alpha }{d}_{x}+\ _{\ _{1}y}\overset{\alpha }{d}_{y}$ acts on
fractional differential forms in N--adapted form; the value $\ ^{\alpha }%
\mathbf{d:=}\ ^{\alpha }\mathbf{e}^{\beta }\ ^{\alpha }\mathbf{e}_{\beta }$
splits into exterior h- and v--derivatives when
\begin{equation*}
\ _{\ _{1}x}\overset{\alpha }{d}_{x}:=(dx^{i})^{\alpha }\ \ _{\ _{1}x}%
\overset{\alpha }{\underline{\partial }}_{i}=\ ^{\alpha }e^{j}\ ^{\alpha }%
\mathbf{e}_{j}\mbox{ and }_{\ _{1}y}\overset{\alpha }{d}_{y}:=(dy^{a})^{%
\alpha }\ \ _{\ _{1}x}\overset{\alpha }{\underline{\partial }}_{a}=\
^{\alpha }\mathbf{e}^{b}\ ^{\alpha }e_{b}.
\end{equation*}

The torsion and curvature of a fractional d--connection $\overset{\alpha }{%
\mathbf{D}}=\{\ ^{\alpha }\mathbf{\Gamma }_{\ \beta \gamma }^{\tau }\}$ can
be defined and computed, respectively, as fractional 2--forms,
\begin{eqnarray}
\ ^{\alpha }\mathcal{T}^{\tau } &\doteqdot &\overset{\alpha }{\mathbf{D}}\
^{\alpha }\mathbf{e}^{\tau }=\ ^{\alpha }\mathbf{d}\ ^{\alpha }\mathbf{e}%
^{\tau }+\ ^{\alpha }\mathbf{\Gamma }_{\ \beta }^{\tau }\wedge \ ^{\alpha }%
\mathbf{e}^{\beta }\mbox{ and }  \label{tors} \\
\ ^{\alpha }\mathcal{R}_{~\beta }^{\tau } &\doteqdot &\overset{\alpha }{%
\mathbf{D}}\mathbf{\ ^{\alpha }\Gamma }_{\ \beta }^{\tau }=\ ^{\alpha }%
\mathbf{d\ ^{\alpha }\Gamma }_{\ \beta }^{\tau }-\ ^{\alpha }\mathbf{\Gamma }%
_{\ \beta }^{\gamma }\wedge \ ^{\alpha }\mathbf{\Gamma }_{\ \gamma }^{\tau
}=\ ^{\alpha }\mathbf{R}_{\ \beta \gamma \delta }^{\tau }\ ^{\alpha }\mathbf{%
e}^{\gamma }\wedge \ ^{\alpha }\mathbf{e}^{\delta }.  \notag
\end{eqnarray}

There are two another important geometric objects: the fractional Ricci
tensor $\ ^{\alpha }\mathcal{R}ic=\{\ ^{\alpha }\mathbf{R}_{\alpha \beta
}\doteqdot \ ^{\alpha }\mathbf{R}_{\ \alpha \beta \tau }^{\tau }\}$ with
components
\begin{equation}
\ ^{\alpha }R_{ij}\doteqdot \ ^{\alpha }R_{\ ijk}^{k},\ \ \ ^{\alpha
}R_{ia}\doteqdot -\ ^{\alpha }R_{\ ika}^{k},\ \ ^{\alpha }R_{ai}\doteqdot \
^{\alpha }R_{\ aib}^{b},\ \ ^{\alpha }R_{ab}\doteqdot \ ^{\alpha }R_{\
abc}^{c}  \label{dricci}
\end{equation}%
and the scalar curvature of fractional d--connection $\overset{\alpha }{%
\mathbf{D}},$
\begin{equation}
\ _{s}^{\alpha }\mathbf{R}\doteqdot \ ^{\alpha }\mathbf{g}^{\tau \beta }\
^{\alpha }\mathbf{R}_{\tau \beta }=\ ^{\alpha }R+\ ^{\alpha }S,\ ^{\alpha
}R=\ ^{\alpha }g^{ij}\ ^{\alpha }R_{ij},\ \ ^{\alpha }S=\ ^{\alpha }g^{ab}\
^{\alpha }R_{ab},  \label{sdccurv}
\end{equation}%
with $\ ^{\alpha }\mathbf{g}^{\tau \beta }$ being the inverse coefficients
to a d--metric (\ref{m1}).

We can introduce the Einstein tensor $\ ^{\alpha }\mathcal{E}ns=\{\ ^{\alpha
}\mathbf{G}_{\alpha \beta }\},$
\begin{equation}
\ ^{\alpha }\mathbf{G}_{\alpha \beta }:=\ ^{\alpha }\mathbf{R}_{\alpha \beta
}-\frac{1}{2}\ ^{\alpha }\mathbf{g}_{\alpha \beta }\ \ _{s}^{\alpha }\mathbf{%
R.}  \label{enstdt}
\end{equation}

For various applications, we can consider more special classes of
d--connections:

\begin{itemize}
\item There is a unique canonical metric compatible fractional d--connection
$\ ^{\alpha }\widehat{\mathbf{D}}=\{\ ^{\alpha }\widehat{\mathbf{\Gamma }}%
_{\ \alpha \beta }^{\gamma }=\left( \ ^{\alpha }\widehat{L}_{jk}^{i},\
^{\alpha }\widehat{L}_{bk}^{a},\ ^{\alpha }\widehat{C}_{jc}^{i},\ ^{\alpha }%
\widehat{C}_{bc}^{a}\right) \},$ when $\ ^{\alpha }\widehat{\mathbf{D}}\
\left( \ ^{\alpha }\mathbf{g}\right) =0,$ satisfying the conditions $\
^{\alpha }\widehat{T}_{\ jk}^{i}=0$ and $\ ^{\alpha }\widehat{T}_{\
bc}^{a}=0,$ but $\ ^{\alpha }\widehat{T}_{\ ja}^{i},\ ^{\alpha }\widehat{T}%
_{\ ji}^{a}$ and $\ ^{\alpha }\widehat{T}_{\ bi}^{a}$ are not zero. The
N--adapted coefficients are given in explicitly form in our works \cite%
{vrfrf,vrfrg,bv1,bv2,bv3,bv4}.

\item The fractional Levi--Civita connection $\ ^{\alpha }\nabla =\{\ \
^{\alpha }\Gamma _{\ \alpha \beta }^{\gamma }\}$ can be defined in standard
from but for the fractional Caputo left derivatives acting on the
coefficients of a fractional metric.
\end{itemize}

On spaces with nontrivial nonholonomic structure, it is preferred to work on
$\overset{\alpha }{\mathbf{V}}$ with $\ ^{\alpha }\widehat{\mathbf{D}}=\{\
^{\alpha }\widehat{\mathbf{\Gamma }}_{\ \tau \beta }^{\gamma }\}$ instead of
$\ ^{\alpha }\nabla $ (the last one is not adapted to the N--connection
splitting (\ref{whit})). The torsion $\ ^{\alpha }\widehat{\mathcal{T}}%
^{\tau }$ (\ref{tors}) \ of $\ ^{\alpha }\widehat{\mathbf{D}}$ is uniquely
induced nonholonomically by off--diagonal coefficients of the d--metric (\ref%
{m1}).

With respect to N--adapted fractional bases (\ref{dder}) and (\ref{ddif}),
the coefficients of the fractional Levi--Civita and canonical d--connection
satisfy the distorting relations
\begin{equation}
\ ^{\alpha }\Gamma _{\ \alpha \beta }^{\gamma }=\ ^{\alpha }\widehat{\mathbf{%
\Gamma }}_{\ \alpha \beta }^{\gamma }+\ \ ^{\alpha }Z_{\ \alpha \beta
}^{\gamma },  \label{cdeft}
\end{equation}%
where the N--adapted coefficients of distortion tensor $\ Z_{\ \alpha \beta
}^{\gamma }$ \ are computed in Ref. \cite{bv4}.

An unified approach to Einstein--Lagrange/Finsler gravity for arbitrary
integer and non--integer dimensions is possible for the fractional canonical
d--connection $\ ^{\alpha }\widehat{\mathbf{D}}.$ The fractional
gravitational field equations are formulated for the Einstein d--tensor (\ref%
{enstdt}), following the same principle of constructing the matter source $\
^{\alpha }\mathbf{\Upsilon }_{\beta \delta }$ as in general relativity but
for fractional metrics and d--connections,%
\begin{equation}
\ ^{\alpha }\widehat{\mathbf{E}}_{\ \beta \delta }=\ ^{\alpha }\mathbf{%
\Upsilon }_{\beta \delta }.  \label{fdeq}
\end{equation}%
Such a system of integro--differential equations for generalized connections
can be restricted to fractional nonholonomic configurations for $\ ^{\alpha
}\nabla $ if we impose the additional constraints%
\begin{equation}
\ ^{\alpha }\widehat{L}_{aj}^{c}=\ ^{\alpha }e_{a}(\ ^{\alpha }N_{j}^{c}),\
\ ^{\alpha }\widehat{C}_{jb}^{i}=0,\ \ ^{\alpha }\Omega _{\ ji}^{a}=0.
\label{frconstr}
\end{equation}

There are not theoretical or experimental evidences that for fractional
dimensions we must impose conditions of type (\ref{frconstr}) but they have
certain physical motivation if we develop models which in integer limits
result in the general relativity theory.

\section{Exact Solutions in Fractional Gravity}

We studied in detail \cite{vrfrg} what type of conditions must satisfy the
coefficients of a metric (\ref{m1}) for generating exact solutions of the
fractional Einstein equations (\ref{fdeq}). For simplicity, we can use a
''prime'' dimension splitting of type $2+2$ when coordinated are labeled in
the form $u^{\beta }=(x^{j},y^{3}=v,y^{4}),$ for $i,j,...=1,2.$ and the
metric ansatz has one Killing symmetry when the coefficients do not depend
explicitly on variable $y^{4}.$

\subsection{Separation of equations for fractional and integer dimensions}

The solutions of equations can be constructed for a general source of type%
\footnote{%
such parametrizations of energy--momentum tensors are quite general ones for
\ various types of matter sources}
\begin{equation*}
\ ^{\alpha }\Upsilon _{\ \ \beta }^{\alpha }=diag[\mathbf{\ }^{\alpha
}\Upsilon _{\gamma };\mathbf{\ }^{\alpha }\Upsilon _{1}=\mathbf{\ }^{\alpha
}\Upsilon _{2}=\mathbf{\ }^{\alpha }\Upsilon _{2}(x^{k},v);\mathbf{\ }%
^{\alpha }\Upsilon _{3}=\mathbf{\ }^{\alpha }\Upsilon _{4}=\mathbf{\ }%
^{\alpha }\Upsilon _{4}(x^{k})]
\end{equation*}%
For such sources and ansatz with Killing symmetries for metrics, the
Einstein equations (\ref{fdeq}) can be integrated in general form.

We can construct 'non--Killing' solutions depending on all coordinates when
\begin{eqnarray}
\mathbf{\ }^{\alpha }\mathbf{g} &\mathbf{=}&\mathbf{\ }^{\alpha }g_{i}(x^{k})%
\mathbf{\ }^{\alpha }{dx^{i}\otimes \mathbf{\ }^{\alpha }dx^{i}}+\mathbf{\ }%
^{\alpha }\omega ^{2}(x^{j},v,y^{4})\mathbf{\ }^{\alpha }h_{a}(x^{k},v)%
\mathbf{\ }^{\alpha }\mathbf{e}^{a}{\otimes }\mathbf{\ }^{\alpha }\mathbf{e}%
^{a},  \notag \\
\mathbf{\ }^{\alpha }\mathbf{e}^{3} &=&\mathbf{\ }^{\alpha }dy^{3}+\mathbf{\
}^{\alpha }w_{i}(x^{k},v)\ ^{\alpha }dx^{i},\mathbf{\ }^{\alpha }\mathbf{e}%
^{4}=\mathbf{\ }^{\alpha }dy^{4}+\mathbf{\ }^{\alpha }n_{i}(x^{k},v)\mathbf{%
\ }^{\alpha }dx^{i},  \label{ansgensol}
\end{eqnarray}%
for any $\mathbf{\ }^{\alpha }\omega $ for which
\begin{equation*}
\mathbf{\ }^{\alpha }\mathbf{e}_{k}\mathbf{\ }^{\alpha }\omega =\overset{%
\alpha }{\underline{\partial }}_{k}\mathbf{\ }^{\alpha }\omega +\mathbf{\ }%
^{\alpha }w_{k}\mathbf{\ }^{\alpha }\omega ^{\ast }+\mathbf{\ }^{\alpha
}n_{k}\overset{\alpha }{\underline{\partial }}_{y^{4}}\mathbf{\ }^{\alpha
}\omega =0.
\end{equation*}%
Configurations with fractional Levi--Civita connection $\mathbf{\ }^{\alpha
}\nabla ,$ of type (\ref{frconstr}), can be extracted by imposing additional
constraints
\begin{eqnarray}
\mathbf{\ }^{\alpha }w_{i}^{\ast } &=&\mathbf{\ }^{\alpha }\mathbf{e}_{i}\ln
|\mathbf{\ }^{\alpha }h_{4}|,\mathbf{\ }^{\alpha }\mathbf{e}_{k}\mathbf{\ }%
^{\alpha }w_{i}=\mathbf{\ }^{\alpha }\mathbf{e}_{i}\mathbf{\ }^{\alpha
}w_{k},\ \   \notag \\
\mathbf{\ }^{\alpha }n_{i}^{\ast } &=&0,\ \overset{\alpha }{\underline{%
\partial }}_{i}\mathbf{\ }^{\alpha }n_{k}=\overset{\alpha }{\underline{%
\partial }}_{k}\mathbf{\ }^{\alpha }n_{i},  \label{frconstr1}
\end{eqnarray}%
where the partial derivatives are
\begin{equation*}
\mathbf{\ }^{\alpha }a^{\bullet }=\overset{\alpha }{\underline{\partial }}%
_{1}a=_{\ _{1}x^{1}}\overset{\alpha }{\underline{\partial }}%
_{x^{1}}{}^{\alpha }a,\ \mathbf{\ }^{\alpha }a^{\prime }=\overset{\alpha }{%
\underline{\partial }}_{2}a=_{\ _{1}x^{2}}\overset{\alpha }{\underline{%
\partial }}_{x^{2}}{}^{\alpha }a,\ \ \mathbf{\ }^{\alpha }a^{\ast }=\overset{%
\alpha }{\underline{\partial }}_{v}a=_{\ _{1}v}\overset{\alpha }{\underline{%
\partial }}_{v}{}^{\alpha }a,
\end{equation*}%
being used the left Caputo fractional derivatives (\ref{frlcb}).

\subsection{Solutions with $\mathbf{\ }^{\protect\alpha }h_{3,4}^{\ast }\neq
0$ and $\ ^{\protect\alpha }\Upsilon _{2,4}\neq 0$}

For simplicity, we provide only a class of exact solution with metrics of
type (\ref{ansgensol}) when $\mathbf{\ }^{\alpha }h_{3,4}^{\ast }\neq 0$ (in
Ref. \cite{vrfrg}, there are analyzed all possibilities for coefficients%
\footnote{%
by nonholonomic transforms, various classes of solutions can be transformed
from one to another}) We consider the ansatz
\begin{eqnarray}
\ \mathbf{\ }^{\alpha }\mathbf{g} &\mathbf{=}&e^{\mathbf{\ }^{\alpha }\psi
(x^{k})}\mathbf{\ }^{\alpha }{dx^{i}\otimes \mathbf{\ }^{\alpha }dx^{i}}%
+h_{3}(x^{k},v)\mathbf{\ }^{\alpha }\mathbf{e}^{3}{\otimes }\mathbf{\ }%
^{\alpha }\mathbf{e}^{3}+h_{4}(x^{k},v)\mathbf{\ }^{\alpha }\mathbf{e}^{4}{%
\otimes }\mathbf{\ }^{\alpha }\mathbf{e}^{4},  \notag \\
\mathbf{\ }^{\alpha }\mathbf{e}^{3} &=&\mathbf{\ }^{\alpha }dv+\mathbf{\ }%
^{\alpha }w_{i}(x^{k},v)\mathbf{\ }^{\alpha }dx^{i},\mathbf{\ }^{\alpha }%
\mathbf{e}^{4}=\mathbf{\ }^{\alpha }dy^{4}+\mathbf{\ }^{\alpha
}n_{i}(x^{k},v)\mathbf{\ }^{\alpha }dx^{i}  \label{genans}
\end{eqnarray}

We consider any nonconstant $\mathbf{\ }^{\alpha }\phi =\mathbf{\ }^{\alpha
}\phi (x^{i},v)$ as a generating function. We have to solve respectively the
two dimensional fractional Laplace equation, for $\ \mathbf{\ }^{\alpha
}g_{1}=\ \mathbf{\ }^{\alpha }g_{2}=e^{\ \mathbf{\ }^{\alpha }\psi (x^{k})}.$
Then we integrate on $v,$ in order to determine $\mathbf{\ }^{\alpha }h_{3},$
$\mathbf{\ }^{\alpha }h_{4}$ and $\mathbf{\ }^{\alpha }n_{i},$ and solve
algebraic equations, for $\mathbf{\ }^{\alpha }w_{i}.$ We obtain (computing
consequently for a chosen $\mathbf{\ }^{\alpha }\phi (x^{k},v)$)
\begin{eqnarray}
\mathbf{\ }^{\alpha }g_{1} &=&\mathbf{\ }^{\alpha }g_{2}=e^{\mathbf{\ }%
^{\alpha }\psi (x^{k})},\mathbf{\ }^{\alpha }h_{3}=\pm \ \frac{|\mathbf{\ }%
^{\alpha }\phi ^{\ast }(x^{i},v)|}{\mathbf{\ }^{\alpha }\Upsilon _{2}},\
\label{gsol1} \\
\mathbf{\ }^{\alpha }h_{4} &=&\ \mathbf{\ }_{0}^{\alpha }h_{4}(x^{k})\pm \
2_{\ _{1}v}\overset{\alpha }{I}_{v}\frac{(\exp [2\ \mathbf{\ }^{\alpha }\phi
(x^{k},v)])^{\ast }}{\mathbf{\ }^{\alpha }\Upsilon _{2}},\   \notag \\
\mathbf{\ }^{\alpha }w_{i} &=&-\overset{\alpha }{\underline{\partial }}_{i}%
\ ^{\alpha }\phi /\mathbf{\ }^{\alpha }\phi ^{\ast },\  
\ ^{\alpha }n_{i} =\ _{1}^{\alpha }n_{k}\left( x^{i}\right) +\
_{2}^{\alpha }n_{k}\left( x^{i}\right) _{\ _{1}v}\overset{\alpha }{I}_{v}[%
\mathbf{\ }^{\alpha }h_{3}/(\sqrt{|\mathbf{\ }^{\alpha }h_{4}|})^{3}],
\notag
\end{eqnarray}%
where $\ \mathbf{\ }_{0}^{\alpha }h_{4}(x^{k}),\ \mathbf{\ }_{1}^{\alpha
}n_{k}\left( x^{i}\right) $ and $\ \mathbf{\ }_{2}^{\alpha }n_{k}\left(
x^{i}\right) $ are integration functions, and $_{\ _{1}v}\overset{\alpha }{I}%
_{v}$ is the fractional integral on variables $v$ and
\begin{eqnarray}
\mathbf{\ }^{\alpha }\phi &=&\ln |\frac{\mathbf{\ }^{\alpha
}h_{4}^{\ast }}{\sqrt{|\mathbf{\ }^{\alpha }h_{3}\mathbf{\ }^{\alpha }h_{4}|}%
}|,\ \mathbf{\ }^{\alpha }\gamma =\left( \ln |\mathbf{\ }^{\alpha
}h_{4}|^{3/2}/|\mathbf{\ }^{\alpha }h_{3}|\right) ^{\ast },  \label{auxphi}
\\
\mathbf{\ }^{\alpha }\alpha _{i} &=&\mathbf{\ }^{\alpha }h_{4}^{\ast }%
\overset{\alpha }{\underline{\partial }}_{k}\ ^{\alpha }\phi ,\ \mathbf{\ }%
^{\alpha }\beta =\mathbf{\ }^{\alpha }h_{4}^{\ast }\ \mathbf{\ }^{\alpha
}\phi ^{\ast }\ .  \notag
\end{eqnarray}%
For $\mathbf{\ }^{\alpha }h_{4}^{\ast }\neq 0;\mathbf{\ }^{\alpha }\Upsilon
_{2}\neq 0,$ we have $\ ^{\alpha }\phi ^{\ast }\neq 0.$ \ The exponent $e^{%
\mathbf{\ }^{\alpha }\psi (x^{k})}$ is the fractional analog of the
''integer'' exponential functions and called the Mittag--Leffer function $%
E_{\alpha }[(x-\ ^{1}x)^{\alpha }].$ For $^{\alpha }\psi (x)=E_{\alpha
}[(x-\ ^{1}x)^{\alpha }],$ we have $\overset{\alpha }{\underline{\partial }}%
_{i}E_{\alpha }=E_{\alpha }.$

We have to constrain the coefficients (\ref{gsol1}) to satisfy the
conditions (\ref{frconstr1}) in order to construct exact solutions for the
Levi--Civita connection $\mathbf{\ }^{\alpha }\nabla .$ To select such
classes of solutions, we can fix a nonholonomic distribution when $\
\mathbf{\ }_{2}^{\alpha }n_{k}\left( x^{i}\right) $ $=0$ and $\ _{1}^{\alpha
}n_{k}\left( x^{i}\right) $ are any functions satisfying the conditions $\
\overset{\alpha }{\underline{\partial }}_{i}\mathbf{\ }_{1}^{\alpha
}n_{k}\left( xj\right) =\overset{\alpha }{\underline{\partial }}_{k}\mathbf{%
\ }_{1}^{\alpha }n_{i}\left( x^{j}\right) .$ The constraints on $\mathbf{\ }%
^{\alpha }\phi (x^{k},v)$ are related to the N--connection coefficients $%
\mathbf{\ }^{\alpha }w_{i}=-\overset{\alpha }{\underline{\partial }}_{i}%
\mathbf{\ }^{\alpha }\phi /\ \mathbf{\ }^{\alpha }\phi ^{\ast }$ following
relations
\begin{eqnarray*}
\left( \mathbf{\ }^{\alpha }w_{i}[\mathbf{\ }^{\alpha }\phi ]\right) ^{\ast
}+\mathbf{\ }^{\alpha }w_{i}[\mathbf{\ }^{\alpha }\phi ]\left( \mathbf{\ }%
^{\alpha }h_{4}[\mathbf{\ }^{\alpha }\phi ]\right) ^{\ast }+\overset{\alpha }%
{\underline{\partial }}_{i}\mathbf{\ }^{\alpha }h_{4}[\mathbf{\ }^{\alpha
}\phi ]=0, &&  \notag \\
\overset{\alpha }{\underline{\partial }}_{i}\mathbf{\ }^{\alpha }w_{k}[%
\mathbf{\ }^{\alpha }\phi ]=\overset{\alpha }{\underline{\partial }}_{k}\
\mathbf{\ }^{\alpha }w_{i}[\mathbf{\ }^{\alpha }\phi ], &&  \label{auxc1}
\end{eqnarray*}%
where, for instance, we denoted by $\mathbf{\ }^{\alpha }h_{4}[\mathbf{\ }%
^{\alpha }\phi ]$ the functional dependence on $\mathbf{\ }^{\alpha }\phi .$
Such conditions are always satisfied for $\mathbf{\ }^{\alpha }\phi =\mathbf{%
\ }^{\alpha }\phi (v)$ or if $\mathbf{\ }^{\alpha }\phi =const$ \ when $%
\mathbf{\ }^{\alpha }w_{i}(x^{k},v)$ can be any functions  with zero $\mathbf{\ }^{\alpha }\beta $ and $\mathbf{\ }%
^{\alpha }\alpha _{i},$ see (\ref{auxphi})).

\section{The Main Theorem on Fractional Solitonic Hierarchies}

In Ref. \cite{bv4,vacap,vanco}, we proved that the geometric data for any
fractional metric (in a model of fractional gravity or geometric mechanics)
naturally define a N--adapted fractional bi--Hamiltonian flow hierarchy
inducing anholonomic fractional solitonic configurations.

\begin{theorem}
\label{mt} For any N--anholonomic fractional manifold with prescribed
fractional d--metric structure, there is a hierarchy of bi-Hamiltonian
N--adapted fractional flows of curves $\gamma (\tau ,\mathbf{l})=h\gamma
(\tau ,\mathbf{l})+v\gamma (\tau ,\mathbf{l})$ described by geometric
nonholonomic fractional map equations. The $0$ fractional flows are defined
as convective (traveling wave) maps%
\begin{equation*}
\gamma _{\tau }=\gamma _{\mathbf{l}},\mbox{\ distinguished \ }\left( h\gamma
\right) _{\tau }=\left( h\gamma \right) _{h\mathbf{X}}\mbox{\ and \ }\left(
v\gamma \right) _{\tau }=\left( v\gamma \right) _{v\mathbf{X}}.
\label{trmap}
\end{equation*}%
There are \ fractional +1 flows defined as non--stretching mKdV maps%
\begin{eqnarray*}
-\left( h\gamma \right) _{\tau } &=&\ ^{\alpha }\mathbf{D}_{h\mathbf{X}%
}^{2}\left( h\gamma \right) _{h\mathbf{X}}+\frac{3}{2}\left| \ ^{\alpha }%
\mathbf{D}_{h\mathbf{X}}\left( h\gamma \right) _{h\mathbf{X}}\right| _{h%
\mathbf{g}}^{2}~\left( h\gamma \right) _{h\mathbf{X}},  \label{1map} \\
-\left( v\gamma \right) _{\tau } &=&\ ^{\alpha }\mathbf{D}_{v\mathbf{X}%
}^{2}\left( v\gamma \right) _{v\mathbf{X}}+\frac{3}{2}\left| \ ^{\alpha }%
\mathbf{D}_{v\mathbf{X}}\left( v\gamma \right) _{v\mathbf{X}}\right| _{v%
\mathbf{g}}^{2}~\left( v\gamma \right) _{v\mathbf{X}},  \notag
\end{eqnarray*}%
and fractional +2,... flows as higher order analogs. Finally, the fractional
-1 flows are defined by the kernels of recursion fractional operators
 inducing non--stretching fractional maps%
$\ ^{\alpha }\mathbf{D}_{h\mathbf{Y}}\left( h\gamma \right) _{h\mathbf{X}}=0$
 and $\ ^{\alpha }\mathbf{D}_{v\mathbf{Y}}\left( v\gamma \right) _{v\mathbf{X}}=0.$
\end{theorem}

\setcounter{equation}{0} \renewcommand{\theequation}
{A.\arabic{equation}} \setcounter{subsection}{0}
\renewcommand{\thesubsection}
{A.\arabic{subsection}}

\appendix

\section{Fractional Caputo N--anholonomic Manifolds}

\label{sappa}

The fractional left, respectively, right Caputo derivatives are defined by
formulas
\begin{eqnarray}
&&\ _{\ _{1}x}\overset{\alpha }{\underline{\partial }}_{x}f(x):=\frac{1}{%
\Gamma (s-\alpha )}\int\limits_{\ \ _{1}x}^{x}(x-\ x^{\prime })^{s-\alpha
-1}\left( \frac{\partial }{\partial x^{\prime }}\right) ^{s}f(x^{\prime
})dx^{\prime };  \label{lfcd} \\
&&\ _{\ x}\overset{\alpha }{\underline{\partial }}_{\ _{2}x}f(x):=\frac{1}{%
\Gamma (s-\alpha )}\int\limits_{x}^{\ _{2}x}(x^{\prime }-x)^{s-\alpha
-1}\left( -\frac{\partial }{\partial x^{\prime }}\right) ^{s}f(x^{\prime
})dx^{\prime }\ .  \notag
\end{eqnarray}%
We can introduce $\ \overset{\alpha }{d}:=(dx^{j})^{\alpha }\ \ _{\ 0}%
\overset{\alpha }{\underline{\partial }}_{j}$ for the fractional absolute
differential, where $\ \overset{\alpha }{d}x^{j}=(dx^{j})^{\alpha }\frac{%
(x^{j})^{1-\alpha }}{\Gamma (2-\alpha )}$ if $\ _{1}x^{i}=0.$ Such formulas
allow us to elaborate the concept of fractional tangent bundle $\overset{%
\alpha }{\underline{T}}M,$ \ for $\alpha \in (0,1),$ associated to a
manifold $M$ of necessary smooth class and integer $\dim M=n.$\footnote{%
For simplicity, we may write both the integer and fractional local
coordinates in the form $u^{\beta }=(x^{j},y^{a}).$ We underlined the symbol
$T$ in order to emphasize that we shall associate the approach to a
fractional Caputo derivative.} \

Let us denote by $L_{z}(\ _{1}x,\ _{2}x)$ the set of those Lesbegue
measurable functions $f$ on $[\ _{1}x,\ _{2}x]$ \ when $||f||_{z}=(\int%
\limits_{_{1}x}^{_{2}x}|f(x)|^{z}dx)^{1/z}<\infty $ and $C^{z}[\ _{1}x,\
_{2}x]$ be the space of functions which are $z$ times continuously
differentiable on this interval. For any real--valued function $f(x)$
defined on a closed interval $[\ _{1}x,\ _{2}x],$ there is a function $%
F(x)=_{\ _{1}x}\overset{\alpha }{I}_{x}\ f(x)$ defined by the fractional
Riemann--Liouville integral $\ _{\ _{1}x}\overset{\alpha }{I}_{x}f(x):=\frac{%
1}{\Gamma (\alpha )}\int\limits_{_{1}x}^{x}(x-x^{\prime })^{\alpha
-1}f(x^{\prime })dx^{\prime },$ when $f(x)=\ _{\ _{1}x}\overset{\alpha }{%
\underline{\partial }}_{x}F(x),$ for all $x\in \lbrack \ _{1}x,\ _{2}x],$
satisfies the conditions
\begin{eqnarray}
\ _{\ _{1}x}\overset{\alpha }{\underline{\partial }}_{x}\left( _{\ _{1}x}%
\overset{\alpha }{I}_{x}f(x)\right) &=&f(x),\ \alpha >0,  \label{aux01} \\
_{\ _{1}x}\overset{\alpha }{I}_{x}\left( \ _{\ _{1}x}\overset{\alpha }{%
\underline{\partial }}_{x}F(x)\right) &=&F(x)-F(\ _{1}x),\ 0<\alpha <1.
\notag
\end{eqnarray}

We can consider fractional (co) frame bases on $\overset{\alpha }{\underline{%
T}}M.$ For instance, a fractional frame basis $\overset{\alpha }{\underline{e%
}}_{\beta }=e_{\ \beta }^{\beta ^{\prime }}(u^{\beta })\overset{\alpha }{%
\underline{\partial }}_{\beta ^{\prime }}$\ is connected via a vierlbein
transform $e_{\ \beta }^{\beta ^{\prime }}(u^{\beta })$ with a fractional
local coordinate basis
\begin{equation}
\overset{\alpha }{\underline{\partial }}_{\beta ^{\prime }}=\left( \overset{%
\alpha }{\underline{\partial }}_{j^{\prime }}=_{\ _{1}x^{j^{\prime }}}%
\overset{\alpha }{\underline{\partial }}_{j^{\prime }},\overset{\alpha }{%
\underline{\partial }}_{b^{\prime }}=_{\ _{1}y^{b^{\prime }}}\overset{\alpha
}{\underline{\partial }}_{b^{\prime }}\right) ,  \label{frlcb}
\end{equation}%
for $j^{\prime }=1,2,...,n$ and $b^{\prime }=n+1,n+2,...,n+n.$ The
fractional co--bases are related via $\overset{\alpha }{\underline{e}}^{\
\beta }=e_{\beta ^{\prime }\ }^{\ \beta }(u^{\beta })\overset{\alpha }{d}%
u^{\beta ^{\prime }},$ where
\begin{equation}
\ _{\ }\overset{\alpha }{d}u^{\beta ^{\prime }}=\left( (dx^{i^{\prime
}})^{\alpha },(dy^{a^{\prime }})^{\alpha }\right) .  \label{frlccb}
\end{equation}

The fractional absolute differential $\overset{\alpha }{d}$ is written in
the form
\begin{equation*}
\overset{\alpha }{d}:=(dx^{j})^{\alpha }\ \ _{\ 0}\overset{\alpha }{%
\underline{\partial }}_{j},\mbox{ where }\ \overset{\alpha }{d}%
x^{j}=(dx^{j})^{\alpha }\frac{(x^{j})^{1-\alpha }}{\Gamma (2-\alpha )},
\end{equation*}%
where we consider $\ _{1}x^{i}=0.$

\end{document}